\documentclass[a4paper,fleqn,usenatbib]{mnras}

\usepackage[T1]{fontenc}
\usepackage{ae,aecompl}
\usepackage{graphicx}	% Including figure files
\usepackage{amsmath}	% Advanced maths commands
\usepackage{amssymb}	% Extra maths symbols

\title[The collisionless NTSI]{The interplay of the collisionless nonlinear thin-shell instability with the ion acoustic instability}

\author[M. E. Dieckmann et al.]{
M. E. Dieckmann$^{1}$\thanks{E-mail: mark.e.dieckmann@liu.se}
D. Folini$^{2}$
R. Walder$^{2}$
\\
$^{1}$Department of Science and Technology (ITN), Link\"oping University, 60174 Norrk\"oping, Sweden.\\
$^{2}$Universit\'{e} de Lyon 1, ENS de Lyon, CNRS, Centre de Recherche Astrophysique de Lyon UMR5574, F-69007, Lyon, France
}
\date{}

\begin{document}
\label{firstpage}
\pagerange{\pageref{firstpage}--\pageref{lastpage}}
\maketitle

\begin{abstract}

The nonlinear thin-shell instability (NTSI) may explain some of the turbulent hydrodynamic structures that are observed close to the collision boundary of energetic astrophysical outflows. It develops in nonplanar shells that are bounded on either side by a hydrodynamic shock, provided that the amplitude of the seed oscillations is sufficiently large. The hydrodynamic NTSI has a microscopic counterpart in collisionless plasma. A sinusoidal displacement of a thin shell, which is formed by the collision of two clouds of unmagnetized electrons and protons, grows and saturates on timescales of the order of the inverse proton plasma frequency. Here we increase the wavelength of the seed perturbation by a factor 4 compared to that in a previous study. Like in the case of the hydrodynamic NTSI, the increase in the wavelength reduces the growth rate of the microscopic NTSI. The prolonged growth time of the microscopic NTSI allows the waves, which are driven by the competing ion acoustic instability, to grow to a large amplitude before the NTSI saturates and they disrupt the latter. The ion acoustic instability thus imposes a limit on the largest wavelength that can be destabilized by the NTSI in collisionless plasma. The limit can be overcome by binary collisions. We bring forward evidence for an overstability of the collisionless NTSI.
\end{abstract}


\section{Introduction}

The boundary between an energetic large-scale astrophysical outflow and an ambient medium like the interstellar medium (ISM) is prone to a plethora of hydrodynamic instabilities, most notably the Rayleigh-Taylor instability, the Kelvin-Helmholtz instability and thin-shell instabilities.

The Rayleigh-Taylor instability can disrupt the boundary between the ISM and the blast shell of a type Ia supernova \citep{Gamezo03} or of a type II supernova \citep{Chevalier92}. It can also develop at the boundary between a pulsar wind and a supernova blast shell \citep{Blondin91,Porth14}. 

The Kelvin-Helmholtz instability limits the growth of the fingers that develop during the nonlinear stage of the Rayleigh-Taylor instability \citep{Chevalier92} and it might be important for radiation- and cosmic ray generation in the shear boundary layers of jets \citep{Stawarz02}. A recent numerical study of this instability is performed by \citet{Palotti08}. 

Linear thin-shell instabilities can form at the collision boundary between the blast shell of a supernova and the ISM \citep{Vishniac83,vanMarle2,Bouquet1,vanMarle1,Bouquet2,Edens10}. A dense shell forms at the front of the blast shell, where it sweeps up the ISM. Initially only the outer boundary between the thin shell and the ISM is a hydrodynamic shock. The inner boundary between the dense shell and the blast shell material changes into a shock at a later time. The linear thin-shell instability can develop prior to the formation of the reverse shock.

A shell that is bounded by two shocks is linearly stable. \citet{Vishniac94} showed however that such a shell is unstable against a sufficiently strong sinusoidal perturbation of its shape and hence it is called the nonlinear thin-shell instability (NTSI). This instability results in turbulent flow inside the shell \citep{Folini06,Folini14} and may play an important role in the thermalization of colliding winds \citep{Walder00}. 

The large time scales over which hydrodynamic astrophysical instabilities develop imply that we can observe only snapshots of their evolution. Some hydrodynamic instabilities can be studied in denser material. A high density of the material compresses the time scale over which the instability evolves and we can observe it from its onset through its nonlinear evolution to its final stage. If we understand the evolution of an instability and know how its density and momentum are distributed at each evolution stage, then we can relate the observed astrophysical gas and plasma structures to the instabilities that created them. Laboratory experiments thus provide essential support for the interpretation of astrophysical observations. 

Laboratory experiments have addressed the Kelvin-Helmholtz instability \citep{Amatucci99} and the Rayleigh-Taylor instability. \citet{Sharp84,Piriz06} provide a description of the Rayleigh-Taylor instability and references to experiments. \citet{Edens10} have observed the linear thin-shell instability at the boundary between a laser-generated blast shell and an ionized ambient medium.

The hydrodynamic \citep{Vishniac94,Blondin96,Lamberts11} and magnetohydrodynamic \citep{Heitsch07,McLeod13} NTSIs have been examined by analytic means and through simulation experiments but, to the best of our knowledge, neither of them has been studied in the laboratory. Its observation in a controlled laboratory experiment would strengthen the case for its existence in astrophysical flows and laboratory studies of its time evolution would shed further light on the topology of the flow patterns it drives.

The basic mechanism of the NTSI can be described as follows. The flow velocity vector of a fluid, which crosses a hydrodynamic shock at an oblique angle, is rotated away from the shock normal because only the velocity component along this normal is decreased by the shock crossing. A fluid flow across a corrugated shock will result in a rotation angle of the velocity vector that is a function of the position along the shock boundary and the inflowing material and the momentum it carries will thus be spatially redistributed in the downstream region. This redistribution amplifies the thin shell's initial corrugation.

% The absence of binary collisions between particles implies that two colliding plasma clouds form an overlap layer, in which the ions of both clouds can move independently for some time. The ion density in this overlap layer is approximately the sum of the ion densities of both clouds and it is thus higher than the density of the surrounding plasma. Thermal diffusion lets electrons stream out of this overlap layer and the ions can not follow instantly due to their larger inertia. A negative net charge develops just outside of the overlap layer and the outflowing electrons leave behind a positive net charge inside the overlap layer. The space charge gives rise to an ambipolar electric field, which points antiparallel to the plasma density gradient. The inflowing ions that make up almost all of the mass density of the plasma are slowed down along the electric field. Their velocity component normal to the electric field is unchanged. Their velocity vector is thus rotated in the same way as that of the fluid at a hydrodynamic shock and we obtain an instability. 

The particle-in-cell (PIC) simulation study by \citet{Dieckmann15} showed that an analogue to the hydrodynamic NTSI exists in a collisionless plasma. The velocity vector of the ions that flow into the shell is rotated by the ambipolar electric field, which is antiparallel to the density gradient at the shell's boundaries. 

Here we examine in more detail the evolution and the saturation of the NTSI in collisionless plasma by means of a particle-in-cell (PIC) simulation. The purpose is to determine if it can develop on a larger scale and for stronger electrostatic shocks than in the simulation by \citet{Dieckmann15}. A broad range of unstable wavelengths and shock strengths would imply that this instability can grow for a wide range of initial conditions, which is a prerequisite for it to be astrophysically relevant and detectable in laboratory plasma. A coupling of the shell's perturbations from the small collisionless scale to larger collisional scales would also imply that the rapidly growing collisionless NTSI could provide the strong seed perturbations that let its large-scale collisional counterpart grow.

Our paper is structured as follows. Section 2 summarizes the PIC simulation method and the initial conditions that we have used for the simulation. Section 2 also describes the double layers and electrostatic shocks \citep{Hershkowitz81} that enclose the thin shell in the collisionless plasma and it summarizes related experimental studies. Section 3 presents our simulation results and we discuss them in Section 4.

\section{Background}

\subsection{The particle-in-cell simulation principle}

Particle-in-cell (PIC) simulation codes are based on the kinetic theory of plasma. The ensemble of the plasma particles that belong to the species $i$ is represented by a phase space density distribution $f_i(\mathbf{x},\mathbf{v},t)$, where $\mathbf{x}$ and $\mathbf{v}$ are the position and velocity coordinates and $t$ is the time. We do not take into account binary collisions in our simulation. The plasma evolution is determined exclusively via the collective electromagnetic fields and $\mathbf{x}$ and $\mathbf{v}$ are thus independent coordinates. The phase space density distribution describes charged particles and its time-evolution is determined by external or self-generated electromagnetic fields, which we compute by Amp\`ere's law and by Faraday's law
\begin{equation}
\mu_0 \epsilon_0 \frac{\partial \mathbf{E}}{\partial t} = \nabla \times \mathbf{B} - \mu_0 \mathbf{J},
\label{AmperesLaw}
\end{equation}
\begin{equation}
\frac{\partial \mathbf{B}}{\partial t} = -\nabla \times \mathbf{E}.
\label{FaradaysLaw}
\end{equation}
The electromagnetic PIC code EPOCH \citep{Arber15} we use solves Eqns. \ref{AmperesLaw} and \ref{FaradaysLaw} on a numerical grid. The time step is $\Delta_t$. It fulfills $\nabla \cdot \mathbf{E} = \rho / \epsilon_0$ and $\nabla \cdot \mathbf{B} = 0$ as constraints. 

Maxwell's equations require the knowledge of the current density $\mathbf{J}$ and of the charge density $\rho$ of the plasma. The phase space density distribution of each plasma species is evolved separately. We obtain the charge contribution of species $i$ from the zero'th moment of its phase space density distribution $\rho_i = q_i \int \, f_i(\mathbf{x}, \mathbf{v}, t) \, d\mathbf{v}$ and its current contribution from the first moment $\mathbf{J}_i = q_i\int \, \mathbf{v} f_i(\mathbf{x},\mathbf{v}, t) \, d\mathbf{v}$. The total charge- and current densities are $\rho = \sum_i \rho_i$ and $\mathbf{J}=\sum_i \mathbf{J}_i$.

The phase space density distribution $f_i (\mathbf{x},\mathbf{v},t)$ of species $i$ is approximated by an ensemble of computational particles (CPs). The $j$'th CP of species $i$ is characterized by the position $\mathbf{x}_j$ and by the momentum $\mathbf{p}_j$. The electromagnetic fields are interpolated from the grid to the position of each CP and a suitably discretized form of Eqn. \ref{LorentzForce} updates its momentum.
\begin{equation}
\frac{d\mathbf{p}_j}{dt} = q_j \left ( \mathbf{E} + \mathbf{v}_j \times \mathbf{B} \right ).
\label{LorentzForce}
\end{equation}
A discretized form of $d\mathbf{x}_j / dt = \mathbf{v}_j$ updates the particle's position. After these updates the current density of each CP is interpolated to the grid, summed up and used to update the electromagnetic fields via Eqns. \ref{AmperesLaw} and \ref{FaradaysLaw}. This cycle is repeated for every time step.

\subsection{Initial conditions}

The plasma, which is composed of protons and electrons with the correct mass ratio $m_p/m_e = 1836$, has initially a constant temperature and density $n_0$ everywhere. The plasma frequency of the electrons is $\omega_{pe} = {(n_0 e^2/m_e \epsilon_0)}^{1/2}$, where $e$ is the elementary charge. The plasma frequency of the protons is $\omega_{pi}=\omega_{pe}/\sqrt{1836}$. The temperatures of the electrons and protons are set to $T_e = $ 1 keV and $T_p = T_e / 5$.

The ion acoustic speed is $c_s={(k_B (\gamma_eT_e + \gamma_pT_p)/m_p)}^{0.5}$, where $m_p$ is the proton mass and $k_B$ the Boltzmann constant. Its value is $c_s = 5 \times 10^5$ m/s if we assume that $\gamma_e = 2$ and $\gamma_p = 3$, which implies 2 degrees of freedom for the electrons and one degree of freedom for the protons. 

The electrons with their low inertia are easily scattered by the thermal fluctuations in the PIC simulation \citep{Dieckmann04b}. The fluctuating electrostatic fields are predominantly polarized in the simulation plane. The scattering of electrons by the electrostatic field fluctuations couples the two velocity components in the simulation plane, which thus has a similar effect as binary collisions \citep{Bret15}, yielding two degrees of freedom for the electrons. 

We express space in units of the electron inertial length $\lambda_s = c/\omega_{pe}$ where $c$ is the speed of light. We resolve the spatial interval $-16.6 \le x \le 16.6$ by 1250 grid cells and the interval $0 \le y \le 6.54$ by 250 grid cells. The boundary conditions along y are periodic, the boundary condition at $x = -16.6$ is open and that at $x=16.6$ is reflecting. The electron species and the proton species are each represented by 250 CPs per cell.

We subdivide the plasma into two clouds, which are initially separated by the boundary $x_B(y) = A_0 \sin{(k_y y)}$ with the wave number $k_y = 2\pi / \lambda_p$. The wave length $\lambda_p = 6.54$ of the seed perturbation equals the box length along y. The amplitude of the seed perturbation is $A_0 = 0.114$ or $A_0 = 0.0175 \lambda_p$. The value of $A_0$ has been selected such that the initial oscillation amplitude is significantly larger than a grid cell while being small compared with $\lambda_p$. 

Each cloud has a mean speed that is spatially uniform. The plasma cloud 1 in the interval $x \le x_B(y)$ has the positive mean speed $v_1 = 1.75 \times 10^6$ m/s equalling $v_1 = 3.5c_s$ along x, while the plasma cloud 2 in the interval $x>x_B(y)$ is initially at rest with $v_2 = 0$. The plasma is initially free of any net charge and current and we set all electromagnetic fields to zero at the simulation's start $t=0$.

\subsection{Collisionless thin shell}\label{sectionD}

\begin{figure}
\includegraphics[width=\columnwidth]{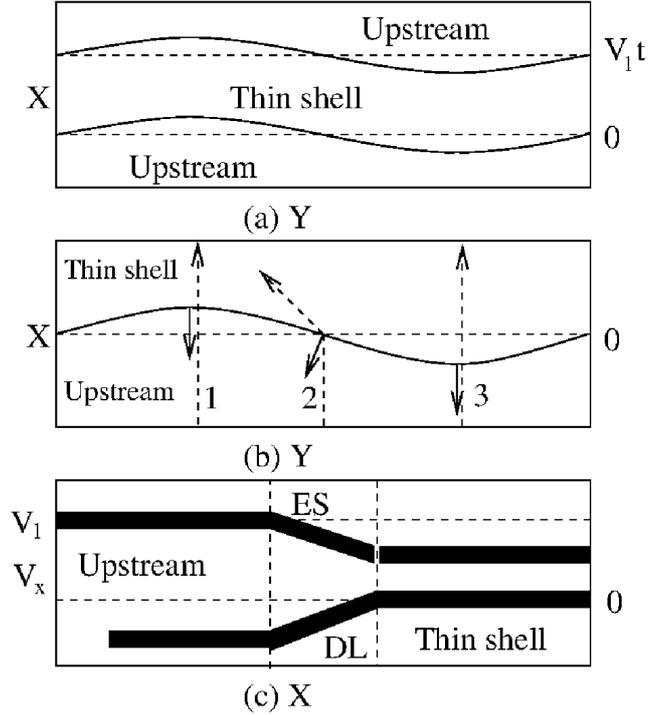}
\caption{Panel (a) illustrates the shape of the thin shell. The lower boundary is determined by the front of the protons that are at rest and the upper one by the protons that move at the speed $v_1$ along $x$. Both boundaries are displaced relative to an average boundary (dashed line). Panel (b) shows the lower boundary, the (solid) electric field vectors and the (dashed) trajectories of protons that move to increasing $x$ at the speed $v_1$. Panel (c) sketches the proton phase space distribution in the $x,v_x$-plane. The dashed vertical lines denote the interval around the lower boundary where we find a nonzero electric field. The abbreviations ES and DL stand for electrostatic shock and double layer.}
\label{Sketch}
\end{figure}

The clouds start to interpenetrate for $t>0$. A thin shell of plasma like that depicted in Fig. \ref{Sketch}(a) forms at the initial contact boundary $x_B(y)$ and expands towards increasing values of $x$ at the speed $v_1$. The density of the plasma in the thin shell is $2n_0$ as long as the protons of both clouds do not interact electromagnetically. We refer to the area covered by the thin shell as the downstream region. A boundary on each side separates the downstream region from the respective plasma cloud. We refer with upstream region to the parts of the plasma cloud that have not yet crossed this boundary. 

Thermal diffusion will lead to a net flow of downstream electrons into the dilute upstream region. A negative charge layer builds up outside of each boundary while the escaping electrons leave behind a positively charged layer just inside of each boundary. A unipolar electrostatic field pulse grows at each boundary of the thin shell between the positive and negative charge layers, which puts the downstream plasma on a higher electric potential than the upstream one. This ambipolar electric field grows and saturates on electron time scales. The field accelerates the protons and it adapts to their changing density distribution.

Figure \ref{Sketch}(b) shows the lower boundary of the thin shell, which remains initially close to $x_B(y)$ because it represents the boundary of the plasma cloud that is at rest. The dashed vectors show the trajectories of three protons that enter the thin shell. Protons 1 and 3 are slowed down as they cross the boundary but their direction is unchanged. Proton 2 is slowed down and deflected by the boundary crossing that decreases only its velocity component along the electric field. Proton 2 is deflected towards an extremal point of $x_B(y)$.

Figure \ref{Sketch}(c) sketches out the proton phase space density distribution in the phase space plane $x,v_x$ parallel to the trajectories of the protons 1 or 3. The vertical dashed lines enclose the spatial interval, in which the electric field is nonzero. The protons of cloud 1, which moves to increasing values of $x$, are found to the left and their mean speed along $x$ is $v_1$. These protons are slowed down by the electric field of the lower boundary when they enter the thin shell. An electrostatic structure that slows down inflowing upstream protons is called electrostatic shock \citep{Forslund70b,Forslund71}. The protons of the stationary cloud 2 are found to the right at a speed $\approx 0$. Their thermal spread implies that some of the protons enter the spatial interval with the nonzero electric field. These protons are accelerated towards the upstream direction and such a structure is called a double layer. \citet{Raadu89} gives a review of double layers in astrophysical plasma. 

Electrostatic shocks and double layers can coexist in a collisionless plasma in the form of a hybrid structure \citep{Hershkowitz81}. We will use this term to denote the nonlinear electrostatic structure that encloses the thin shell unless we discuss its components.

\subsection{The hybrid structure and related experiments}\label{shocksD}

The potential difference between the upstream and the downstream plasma is set by the density jump, which is of the order of $n_0$, and the electron temperature that does usually not vary much across a hybrid structure. If the kinetic energy of the inflowing upstream protons in the shell's rest frame is large compared with the potential energy change at the shell boundary then these protons are hardly slowed down. The colliding clouds will interpenetrate without forming a well-defined dense and localized thin shell. The maximum Mach number of such a shell is thus limited.  Our collision speed $v_1 = 3.5 c_s$ brings us into the regime where the velocity gap between the counterstreaming proton clouds in the shell is initially comparable to $v_1$. A gradually increasing compression of the plasma in the thin shell and the associated growth of the potential difference between the upstream and downstream plasmas reduces in time the gap between the beam velocities \citep{Dieckmann13}. 

Another property of the hybrid structure is that the inflowing upstream protons are not fully thermalized when they enter the downstream region (See Fig. \ref{Sketch}(c)). A thermalization is eventually achieved by the electrostatic turbulence \citep{Dum78,Bale02,Dieckmann14} that is driven by an instability between the counterstreaming proton beams \citep{Forslund70}. In what follows we call it the proton-proton beam instability.

%The upstream protons are slowed down in the shell's reference frame by its positive potential. The piling up of the protons increases the density of the thin shell above the value we would get by adding up the densities of both clouds. The absence of binary collisions allows the protons to traverse the thin shell and they are reaccelerated by the double layer when they reach the opposite side of the thin shell. The more speed the protons loose when they enter the thin shell, the more they are compressed and the more time they need to reach the opposite side of the thin shell thereby enhancing the mass gain of the shell. The velocity loss of the protons is substantial if their kinetic energy in the shell's frame is low, which implies in turn that the upsteam plasma excerts only a low ram pressure on the downstream plasma. 

%A low ram pressure on the downstream plasma and the high thermal pressure in the shell that results from the strong compression lets the shell expand rapidly.  This is true for slow electrostatic shocks and their speed is a large fraction of $c_s$ in the downstream frame. If the expansion speed of the shell is higher than the speed of a proton inside the shell the proton remains trapped, which enhances further the mass accumulation by the shell.

The plasma parameters, which we have selected, are comparable to those in the experiment performed by \citet{Ahmed13}. A thin plasma shell was created in this study by the collision of a blast shell, which was ejected by a laser-ablated solid target, with an ambient medium. The source of the ambient medium was the residual gas, which was contained in the plasma chamber prior to the ablation of the target and got ionized by secondary X-ray emissions from the ablated target. The ultraintense laser pulse and observational time scale that was of the order of 100 picoseconds implied that effects caused by binary collisions between plasma particles were negligible. It may thus be possible to reproduce the NTSI in a collisionless laboratory plasma. 

Binary particle collisions would establish a Maxwellian velocity distribution of the protons in the downstream region. Only few protons are fast enough in such a distribution to catch up with the hybrid structure and be accelerated upstream by its double layer component. Those that make it will collide with the inflowing upstream particles and they will be pushed back to the hybrid structure. As we increase the collisionality of the plasma the hybrid structure will gradually change into a fluid shock. The degree of collisionality in a laboratory plasma experiment depends on the intensity of the laser pulse and on the observational time scale. \citet{Hansen06} observed a collisional shock. 

It is of interest to establish with PIC simulations the range of parameters for which the collisionless NTSI can develop and to test if it can develop in a collisionless laser-plasma experiment. Here we examine if the collisionless NTSI can destabilize a wavelength that exceeds that in \citet{Dieckmann15} by a factor of 4. Further experiments and PIC simulation studies can then examine how the NTSI evolves in collisional plasma.

\section{Simulation results}

We present and discuss the proton density distribution and the distributions of the in-plane electric field and of the out-of-plane magnetic field at several times. In what follows we normalize time as $t = \tilde{t} \omega_{pe}$ where $\tilde{t}$ is expressed in SI units. We select the times $t_1 = 268$, $t_2 = 536$, $t_3 = 1.1 \times 10^3$, $t_4 = 1.6  \times 10^3$, $t_5 = 2.1  \times 10^3$ and $t_6 = 2.7 \times 10^3$. The proton density distribution $n_p$ is normalized to $n_0$, the in-plane electric field $E_p(x,y) = {(E_x^2(x,y) + E_y^2(x,y))}^{1/2}$ is normalized to $m_e \omega_{pe} c / e$ and the out-of-plane magnetic field $B_z (x,y)$ is normalized to $m_e \omega_{pe} / e$. 

The Maxwell equations can be normalized with the aforementioned normalization of the electric and magnetic field if we use $\lambda_s$ to normalize space and $\omega_{pe}^{-1}$ to normalize time (See \citet{Dieckmann08} for details). The Maxwell equations and the particle equations of motion do not depend explicitely on the value of $n_0$ in their normalized form, as long as binary collisions between particles are not important. The value of $n_0$ does not influence in this case the plasma dynamics and $n_0$ only becomes important when we scale the simulation results to SI units. Space and time scale with ${n_0}^{-1/2}$ and the electric and magnetic field amplitudes with ${n_0}^{1/2}$. Table \ref{tableNorm} presents the numerical values of the factors we have to multiply to the positions, times and field amplitudes for several values of $n_0$.

\begin{table}
\caption{The multiplier for the normalized quantities for three values of the electron density $n_0$ expressed in units cm$^{-3}$.}
\label{tableNorm}
\centering
\begin{tabular}{ccccc}
\hline
$n_0$ & $\mathbf{x}$ & $t$ & $\mathbf{E}$ & $\mathbf{B}$ \\ 
$1$ & 5.3 km & 18 $\mu$s & 96.2 V/m & 320 nT \\
$10^3$ & 168 m & 0.56 $\mu$s & 3 kV/m & 10 $\mu$T \\
$10^{14}$ & 0.53 mm & 1.8 ps & 960 MV/m & 3.2 T \\
\hline
\end{tabular}
\end{table}

\begin{figure*}
\centering
   \includegraphics[width=17cm]{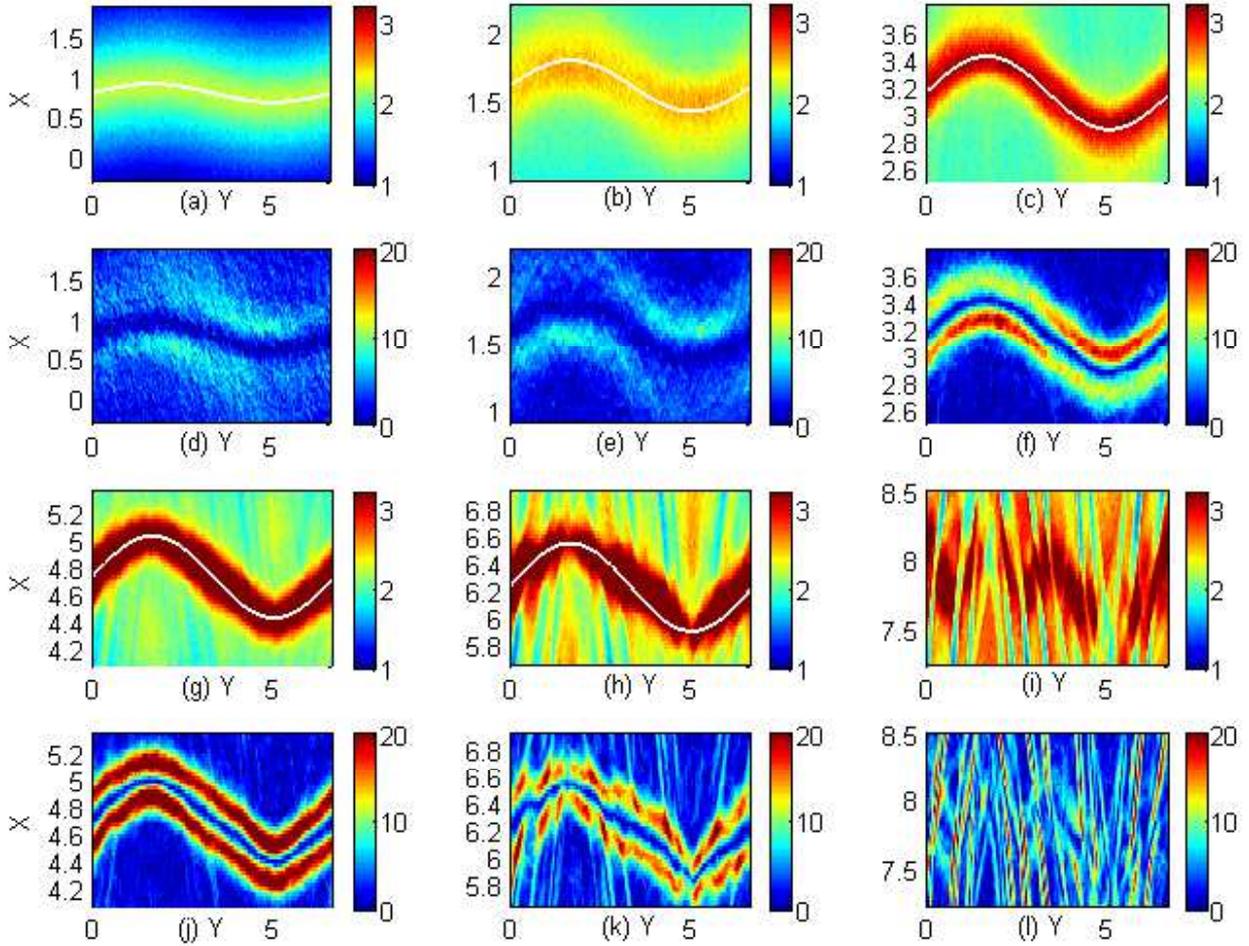}
     \caption{The proton density distribution $n_p (x,y)$ and the in-plane electric field distribution $E_p (x,y)$ multiplied by a factor $10^3$. The first and the third row correspond to $n_p(x,y)$. The second and the fourth row show $E_p (x,y)$. The electric field distribution belonging to a proton density distribution is shown underneath the latter. Panels (a,d) correspond to the time $t_1=268$. Panels (b,e) correspond to $t_2 = 536$. Panels (c,f) correspond to $t_3 = 1.1 \times 10^3$. Panels (g,j) correspond to $t_4 = 1.6 \times 10^3$. Panels (h,k) correspond to $t_5=2.1 \times 10^3$. Panels (i,l) correspond to $t_6 = 2.7 \times 10^3$. A sine wave is fitted to the centre of the thin shell for the times $t_1$ to $t_5$. Table \ref{table1} shows the values of its amplitude and offset along $x$.}
     \label{Bigfigure}
\end{figure*}

Snapshots of $n_p(x,y)$ and $E_p(x,y)$ are displayed in Fig. \ref{Bigfigure}.
The curves $x_i(y) = x_i + A_i \sin{(2 \pi y/ 6.54)}$ are overplotted at the centres of the thin shells at the times $t_i$ with $1 \le i \le 5$. The offset $x_i$ is expressed in units of $\lambda_s$ and the amplitude $A_i$ is normalized to $A_0$. We calculate the normalized speed $v_i = x_i / (t_i v_1)$ and the normalized speed $\Delta A_i = (A_i-A_{i-1})/(v_1\, [t_i - t_{i-1}])$ with which the amplitude grows at the extrema of the oscillation. Table \ref{table1} shows their values.
\begin{table}
\caption{The parameters of the fitted sine curve}
\label{table1}
\centering
\begin{tabular}{llllll}
\hline
Time $t_i$: & $t_1$ & $t_2$ & $t_3$ & $t_4$ & $t_5$ \\
Time value : & 268 & 536 & 1100 & 1600 & 2100 \\
Amplitude $A_i$: & 1.0 & 1.6 & 2.3 & 2.6 & 2.8 \\
Offset $x_i$: & 0.8 & 1.6 & 3.15 & 4.73 & 6.23 \\
Speed $v_i$: & 0.51 & 0.51 & 0.49 & 0.51 & 0.51 \\
Growth speed: $\Delta A_i$: & & 0.19 & 0.21 & 0.1 & 0.07 \\
\hline
\end{tabular}
\end{table}

The normalized speed $v_i \approx v_1/2$ is approximately constant. The centre of the high-density layer of the protons thus moves at the speed $v_1 / 2$ towards increasing values of $x$ as we expect from the global conservation of momentum and the equal cloud densities. The amplitude $A_i$ grows from $t_1$ to $t_3$ at an average value of $0.2 v_1$ or $0.7c_s$. Its growth rate decreases for $t>t_3$. The increase of the amplitude from $A_0$ to $2.8A_0$ demonstrates that the thin shell is unstable against the initial spatial displacement. 

The thickness of the thin shell is about 0.7 at $t_1$ in Fig. \ref{Bigfigure}(a). The positive potential of the thin shell slows down the inflowing upstream plasma. The ensuing pile-up of the protons increases the plasma density within the shell to a value above 2. The proton density has not reached anywhere the value $n_p(x,y)\ge 3$ that we would expect if strong hybrid structures would enclose the thin shell. The potential difference between the thin shell and the surrounding plasma is not yet high enough to reduce significantly the velocity gap in Fig. \ref{Sketch}(c).

The electric field distribution in Fig. \ref{Bigfigure}(d) shows large patches with a low peak amplitude. We do thus not find anywhere large plasma density gradients and, hence, no strong hybrid structure. The electric field amplitude is largest close to the concave boundaries of the thin shell in Fig. \ref{Bigfigure}(a). Both boundaries of the thin shell follow $x_B(y)$. The thin shell has thus merely expanded along x.

Protons have accumulated close to the extrema of the thin shell's oscillation at $y \approx 1.6$ and $y \approx 4.9$ in Fig. \ref{Bigfigure}(b). The density gradient is larger at the concave sections of the thin shell than at the convex sections and it drives a larger ambipolar electric field in Fig. \ref{Bigfigure}(e). The accumulation of protons at the extrema of the thin shell's spatial distribution indicates according to Fig. \ref{Sketch} the onset of the NTSI, which we can understand in the following way. The average speed $v_1/2$ is maintained at the zero-crossings of the thin shell's oscillation at $y=0$ and $y\approx 3.3$ due to an equal density of the colliding proton clouds at these positions. The proton deflection by the thin shell does however increase the number of protons with $v_x \approx v_1$ at $y\approx 1.6$ and it increases the number of protons with $v_x \approx 0$ at $y\approx 4.9$, which alters the momentum balance between both clouds at the extremal points and amplifies the oscillation via a change of the mean speed of the thin shell. Indeed the amplitude of the oscillation has increased to $1.6A_0$. The proton density has increased to a value $n_p \approx 2.5$ in an interval with a width 0.4 along $x$ and the density oscillates along the thin shell with an amplitude of about $0.1$. 

The potential difference between the thin shell and the surrounding plasma increases with the density, which results in a stronger compression. Peak values of $n_p \approx 3.3$ in Fig. \ref{Bigfigure}(c) evidence a strong compression of the upstream plasma when it enters the thin shell. The proton density shows only weak oscillations within the thin shell at this time. The associated electric field distribution $E_p(x,t)$ in Fig. \ref{Bigfigure}(f) demonstrates that the narrow unipolar electric field bands, which are the characteristic of hybrid structures, are strongly modulated along y. Their amplitude peaks at the concave sections, which thus provide the largest density gradients. The amplitude $A_i$ of $x_B(y)$ has grown further to a value 2.3. 

The large density value $n_p \approx 2.4$ seen in Fig. \ref{Bigfigure}(c) at $x\approx 3.6$ and $y\approx 1.6$ can only be explained by an outflow of the protons of the plasma cloud 1, which was collimated by the thin shell. The same is true for the protons of the plasma cloud 2 that are collimated by the thin shell into the region $x\approx 2.6$ and $y \approx 4.6$. The boundaries of the thin shell thus have a double-layer component and the boundaries are indeed hybrid structures. 

Figure \ref{Bigfigure}(g) evidences that the density of the thin shell has equilibrated. The electric field in Fig. \ref{Bigfigure}(j) has a practically constant amplitude along both boundaries and its distribution shows a piecewise linear shape. The electric field of the hybrid structure, which is determined by the density gradient at the shell's boundary, should still deflect most protons towards the extrema of $x_B(y)$. The absent density accumulation at the extrema suggests that a second process is counteracting this mass flow.

The density distribution in Fig. \ref{Bigfigure}(b) is the one expected from Fig. \ref{Sketch}(b). The density distribution has equilibrated sometime between $t_3$ (Fig. \ref{Bigfigure}(c)) and $t_4$ (Fig. \ref{Bigfigure}(g)) and the equilibration time scale $\Delta_t$ is thus between $t_3-t_2$ and $t_4-t_2$ or $500 < \Delta_t < 10^3$. Let us assume that the density oscillates along a planar part of the thin shell, which has a length of $\approx 3 \lambda_s$. The wavelength of the oscillation is thus $k_0 = 2\pi / 3\lambda_s$. The ion acoustic speed is $c_s = 5 \times 10^5$ m/s. One ion acoustic oscillation takes place during $\tilde{t}_s = 2\pi / (k_0 c_s)$, where $\tilde{t}_s$ is given in seconds. We can rewrite this expression as $t_s = \tilde{t}_s \omega_{pe} = 3 c / c_s$, which gives $t_s \approx 1800$. The equilibration we observe thus takes place during about $0.25 < \Delta_t / t_s < 0.5$.  

Ion acoustic waves are charge density waves and such waves can lead to large oscillations of the plasma density. The density equilibration along the shell may thus be tied to such an oscillation. This equilibration coincides with the reduction of $\Delta A_i$ at this time. The amplitude $A_4$ has grown only by $\approx 0.3$ between $t_3$ and $t_4$ and $\Delta A_4 = \Delta A_3 / 2$. This coincidence suggests that the density equilibration is responsible for the decrease of the growth rate, which would imply that the collisionless NTSI is overstable. 

The amplitude growth of the shell's spatial displacement slows down further as we go from $t_4$ to $t_5$ in Fig. \ref{Bigfigure}(h) and we measure the largest value $A_5 \approx 2.8$ of the modulation at this time. The density $n_p(x,y)$ peaks now at $y\approx 3.3$ and $y \approx 0$, which is the opposite of what we expect from the proton deflection by the hybrid structure. This distribution can be explained in terms of an overshoot of the ion acoustic wave, which is further evidence for an oscillation of the proton density distribution along the thin shell. 

The shell remains thin during the entire simulation time and it does thus hardly accumulate material. The slow expansion of the thin shell is favorable for a continuing growth of the oscillation amplitude $A_i$. However, the thin shell starts to break up at the extremal points of the spatial oscillation. The distribution of $n_p(x,y)$ in Fig. \ref{Bigfigure}(h) within the thin shell and that of $E_p(x,y)$ in Fig. \ref{Bigfigure}(k) at its boundaries are both fragmented. The same is true for the proton density distributions in both upstream regions. These density oscillations are the result of a proton-proton beam instability inside the shell and in the upstream region close to it (See Fig. \ref{Sketch}(c)). This instability ultimately seals the fate of the thin shell by giving rise to the growth of strong electrostatic fields with potential variations that are comparable to the potential jump between the thin shell and the inflowing plasma. The destruction of the thin shell by ion acoustic waves is evidenced by Fig. \ref{Bigfigure}(i) and the electric field in Fig. \ref{Bigfigure}(l). 

The mechanism that results in the hydrodynamic NTSI is the transport of material towards the extrema of the thin shell's spatial oscillation. The rotation of the fluid velocity vector by the oblique crossing of a hydrodynamic shock always results in a flow towards the extremal positions, because the fluid is trapped within the thin shell. A hybrid structure can, however, not trap protons within the thin shell. Once the protons reach the opposite side of the thin shell, they are reaccelerated by the double layer and propagate upstream. The thin-shell instability in collisionless plasma is thus only similar to the NTSI if a significant fraction of the protons is indeed moving to the extremal positions of the thin shell at $y \approx 1.6$ and $y \approx 4.9$. We must compare the flow direction of the protons within the shell with the direction of the thin shell. 

\begin{figure} 
\includegraphics[width=\columnwidth]{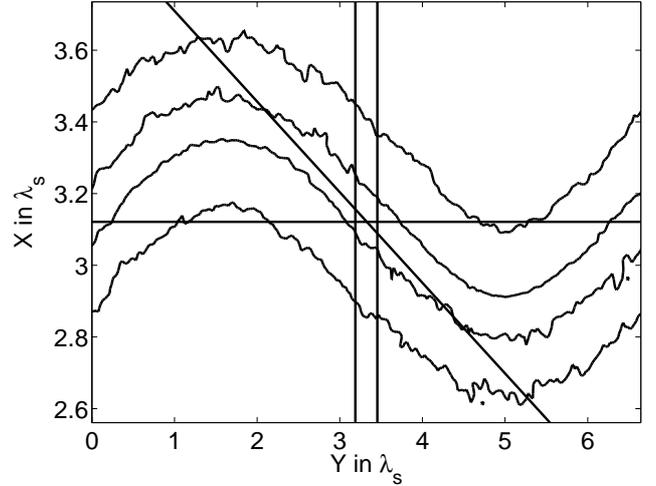}
\caption{The thin shell at the time $t_3$. The contour lines correspond to 0.4 times the peak value of $E_p (x,y)$. The horizontal line denotes $x=x_3$. The two vertical lines delimit the spatial interval from which we will sample the velocities of the computational particles. The diagonal line is oriented at an angle of $20^\circ$ relative to $x=x_3$ at $y \approx 3.3$.}\label{Shellslice}
\end{figure}

We estimate with Fig. \ref{Shellslice} the angle between $x=x_3$ (Table \ref{table1} at the time $t_3$) and the direction of the thin shell at $y\approx 3.3$ to about $20^\circ$. 
Protons that move along this direction in the rest frame of the shell remain inside the shell. 

\begin{figure} 
\includegraphics[width=\columnwidth]{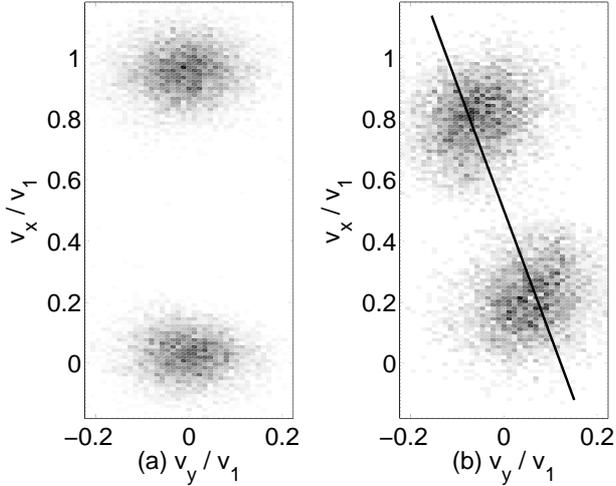}
\caption{The velocity distribution at the time $t_3$ and $y \approx 3.3$ on a linear grayscale and in the reference frame of the simulation box. Panel (a) shows the proton velocity distribution far upstream of the thin shell at low $x$. The beam with $v_x \approx v_1$ corresponds to the protons of cloud 1, which flow towards the thin shell. The lower beam is composed of protons of the cloud 2 that left the thin shell at the opposite side. Panel (b) shows the proton distribution in the centre of the thin shell. The velocity vectors of both beams have been rotated by an angle of approximately $20^\circ$, which is indicated by the overplotted line.}\label{Velocity122}
\end{figure}

We sample the in-plane velocity components $v_x$ and $v_y$ from the protons that are located in the spatial interval, which is delimited by the two vertical lines in Fig. \ref{Shellslice}. The velocity distribution of the protons with $2.5 < x < 2.6$ is shown in Fig. \ref{Velocity122}(a) and that of the protons in the interval $3.1 < x < 3.2$ is shown in Fig. \ref{Velocity122}(b).
The proton distributions are well-separated in the velocity direction both inside and outside of the thin shell. Their relative speed exceeds by far their thermal velocity spread and such a distribution gives rise to the proton-proton beam instability. 

Both proton beams move along the x-direction in Fig. \ref{Velocity122}(a). The beam with $v_x\approx 0$ in  Fig. \ref{Velocity122}(a) consists of protons that crossed the thin shell. The velocity rotation they experience when they enter the thin shell is cancelled out by the rotation in the opposite direction when they leave it.  

The proton velocity vectors are rotated in Fig. \ref{Velocity122}(b) by an angle $\approx 20^\circ$ around the pivot point $v_x=v_1/2$ and $v_y=0$. The velocity distribution inside the thin shell demonstrates that the majority of the protons flow along the thin shell. These protons will eventually reach the extremal positions of the shell's oscillation at $y\approx 1.6$ and $y\approx 4.9$. 

The proton phase space density distribution in the simulation resembles that in the sketch in Fig. \ref{Sketch}(c) if we neglect the proton's lateral velocity component. We do not find any protons that move at the mean speed $v_1/2$ of the shell. The slowdown of the protons by the shell's potential is not sufficiently high to trap them. That would require that the proton speed inside the shell and measured in the shell's rest frame is less than the speed with which the shell's thickness increases. The latter is negligible compared with $v_1$. The protons thus leave the thin shell at the extrema of its oscillation, feeding the collimated outflow seen in Fig. \ref{Bigfigure}(c).

\begin{figure}
\includegraphics[width=\columnwidth]{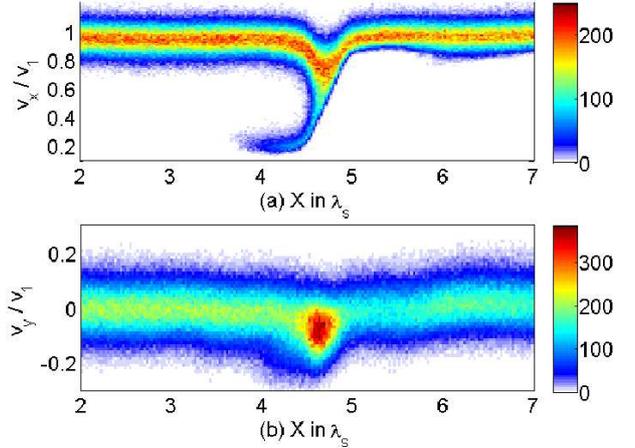}
\caption{Projections of the phase space density distribution of the protons of cloud 1 onto the $x,v_x$ plane (a) and onto the $x,v_y$ plane (b) at $t=t_4$. The colour scale is linear.}
\label{Phasespace183}
\end{figure}

Figure \ref{Phasespace183} shows the phase space density distribution of the protons of cloud 1 averaged over the y-interval, which is delimited by the vertical lines in Fig. \ref{Shellslice}. The thin shell is located at $x\approx 4.7$ (Table \ref{table1}).
The protons that enter the thin shell reach their lowest mean speed $v_x \approx 0.75v_1$ at $x\approx x_4$ in Fig. \ref{Phasespace183}(a) and their mean speed along the y-direction reaches $v_y\approx -0.1 v_1$ in Fig. \ref{Phasespace183}(b). Figure \ref{Phasespace183}(a) shows that the protons are reaccelerated by the double layer at $x\approx 4.9$ when they leave the thin shell and move into the upstream region at $x\approx 5$. Some of the protons are reflected by the electrostatic shock at $x\approx 4.5$ and they form the beam at $x\approx 4$ and $v_x \approx 0.2$. These protons fall behind the thin shell, which moves at the mean speed $v_1/2$ and they thus constitute a shock-reflected proton beam. The proton beam in Fig. \ref{Phasespace183}(a) in the interval $x>5$ is not spatially uniform. The density is lower for $5<x<5.8$ than for $x>5.8$. The growth in time of the thin shell's potential relative to the upstream results in an increasing proton compression within the shell, which reduces temporally the number of protons that exit the thin shell via the double layer.

\begin{figure}
\includegraphics[width=\columnwidth]{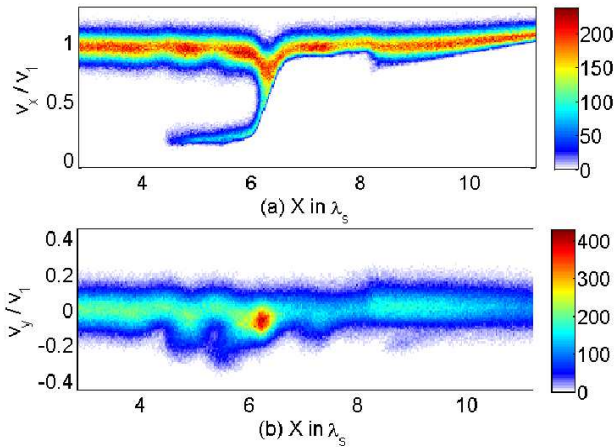}
\caption{Projections of the phase space density distribution of the protons of cloud 1 onto the $x,v_x$ plane (a) and onto the $x,v_y$ plane (b) at $t=t_5$. The colour scale is linear.}
\label{Phasespace241}
\end{figure}

Figure \ref{Phasespace241} shows the projections of the proton phase space density distribution onto the $x,v_x$ plane and onto the $x,v_y$ plane at the time $t=t_5$.
The phase space density distributions are qualitatively similar to those at the previous time but they are more turbulent. The phase space density in the interval $4 < x < 6$ and $v_x \approx v_1$ varies in Fig. \ref{Phasespace241}(a). The density changes are correlated with changes in the mean speed in Fig. \ref{Phasespace241}(b). We attribute these localized changes of the proton's mean speed and density to the ion acoustic waves, which we observe in Fig. \ref{Bigfigure}(h). 

\begin{figure*}
\centering
   \includegraphics[width=17cm]{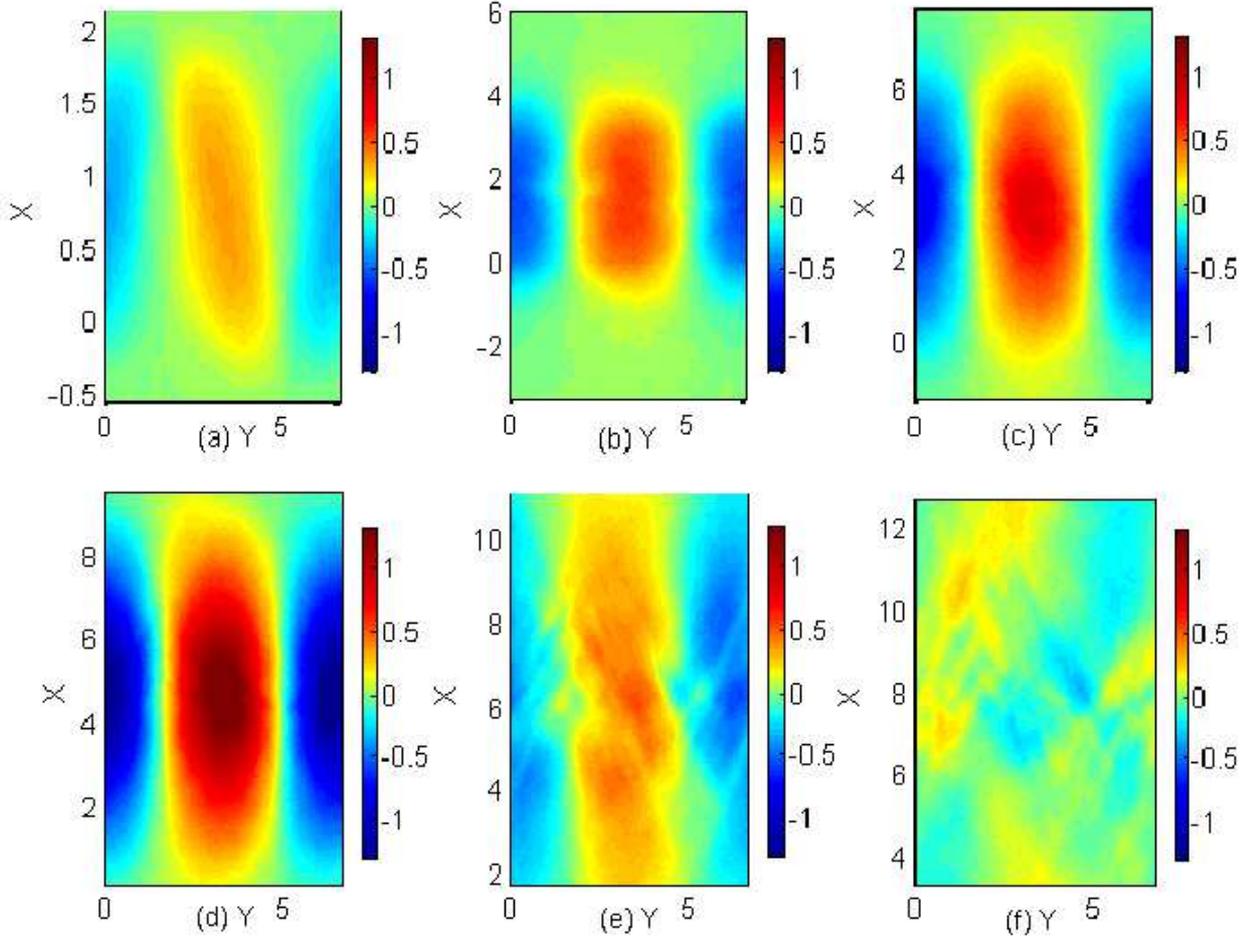}
     \caption{The out-of-plane magnetic field distribution $B_z (x,y)$ multiplied by the factor 100. Panel (a) corresponds to the time $t_1=268$, panel (b) to $t_2 = 536$, panels (c) to $t_3 = 1.1 \times 10^3$, panel (d) to $t_4 = 1.6 \times 10^3$, panels (e) to $t_5=2.1 \times 10^3$ and panels (f) to $t_6 = 2.7 \times 10^3$.} 
     \label{Magneticfigure}
\end{figure*}

According to Fig. \ref{Sketch}(b), the electric field deflects the protons towards the extrema of the shell's oscillation by decelerating them along the normal of the shell's boundary. Electrons that enter the shell should be accelerated along the normal by this field due to their opposite charge. Figure \ref{Magneticfigure} demonstrates that this drift generates magnetic fields.
The magnetic field modulus peaks at $y=0$ and $y=3.3$ and the magnetic field patches are centred around the corresponding value of $x=x_i$. The magnetic field amplitude grows and the magnetic field patches expand until $t=t_4$. 

The potential difference between the shell plasma and the upstream plasma determines the drift velocity between the electrons and protons that enter the shell and, thus, the net current. A growth of this potential difference through an increase of the plasma density within the shell thus results in the growth of the magnetic field energy, as long as the electric fields are well-defined unipolar pulses. 

The magnetic field weakens once the thin shell starts to be fragmented by the ion acoustic instability at $t=t_5$ and all that remains at $t=t_6$ are small-scale magnetic fluctuations. The temporal correlation between the magnetic field collapse and the destruction of the thin shell demonstrates that the latter is the driver of the magnetic field. 

\section{Discussion}

We have examined the collision of two clouds of electrons and protons at a speed that exceeded the ion acoustic speed by a factor 3.5. Their initial contact boundary was sinusoidally displaced along the collision direction. The displacement of the contact boundary resulted in a sinusoidally corrugated thin shell that was formed by the interpenetrating plasma clouds and this corrugation seeded the NTSI. We have confirmed that a wavelength of the seed perturbation, which exceeded that used in the previous simulation study by \citet{Dieckmann15} by a factor of 4, is unstable. A wide range of wavenumbers of the seed perturbation is thus subjected to the NTSI.

We have identified here the proton-proton beam instability as the process that limits the life-time of the thin shell. This instability is known to destroy planar double layers and electrostatic shocks \citep{Karimabadi91,Kato10,Dieckmann15b} and here we have shown that it also affects the nonplanar ones. 

The amplitude of the shell's spatial oscillation grew because the NTSI introduces a spatially varying velocity of the thin shell in the reference frame that moves with the mean speed of the shell. The modulus of the velocity peaked at the extrema of the shell's oscillation and the velocity at these positions reached 70\% of the ion acoustic speed. The amplitude of the thin shell's spatial displacement grew during the simulation to almost three times its initial value before the shell was destroyed by the proton-proton beam instability. 

Our simulation data hints at a possible coupling of the NTSI with ion acoustic oscillations along the thin shell. We have explained the change of the NTSI's growth rate at late times in terms of these oscillations, which would make the collisionless NTSI overstable. Such an overstability has also been observed for the hydrodynamic linear thin-shell instability \citep{Vishniac83}.

The ambipolar electric field at the boundaries of the thin shell deflected the inflowing upstream electrons and protons into different directions. The relative drift of the electrons and the protons resulted in a net current and, thus, in the growth of magnetic fields. 

We can obtain additional qualitative insight into the collisionless NTSI by comparing the simulation results we have obtained here with those discussed in related work. 

The shorter wavelength of the seed perturbation in the simulation by \citet{Dieckmann15} resulted in two important differences. Firstly, the shorter wavelength of the seed perturbation used in that previous work implied that the ratio of the amplitude of the shell's corrugation to the wavelength of the corrugation could grow to a much larger value before the proton-proton beam instability set in. The low maximum ratio that can be reached for long wave lengths of the seed oscillation probably implies that it will be more difficult to observe their growth. Secondly, the larger proton density gradients within the thin shell that developed during the growth phase of the NTSI in the simulation by \citet{Dieckmann15} resulted in ambipolar electric fields along the thin shell that were strong enough to drive nonlinear plasma structures within the thin shell. The lower density gradients within the thin shell that were reached in the present simulation resulted in density oscillations along the thin shell that remained in the linear regime. 

The peak amplitude of the magnetic field strength in the present simulation is four times that in the simulation by \citet{Dieckmann15} and the field patches extended far upstream. The weak magnetic fields observed by \citet{Dieckmann15} remained practically confined to the thin shell. The longer wavelength of seed oscillation we have used here thus generates magnetic fields with a larger energy than those found by \citet{Dieckmann15}.  

The size of the magnetic field patches we found here was comparable to those in the simulation by \citet{MagField15}, where the curved shell was created by a spatial variation of the collision speed. The magnetic field in the present simulation and in that in \citep{MagField15} damped out when the thin shell was destroyed by the proton-proton beam instability, evidencing that the hybrid structures were responsible for its growth.

It is possible to introduce a collision operator into a PIC simulation that emulates the effects of binary collisions between particles. Collisions affect the growth rate of the proton-proton beam instability and if they occur frequently they can thermalize the protons within the thin shell and scatter the proton beam that moves back upstream before an instability sets in. The hybrid structure will probably change into a fluid shock if collisions are frequent. 

The maximum speed, which parts of a hydrodynamic thin shell can reach in the shell's rest frame, is just below the sound speed \citep{Vishniac94}. The sound speed is the hydrodynamic equivalent of the ion acoustic speed in a collisionless plasma, which suggests that we can go from the collisionless to the hydrodynamic limit discussed by \citet{Vishniac94} by increasing the collisionality of the plasma. We will test this hypothesis in future work.

\textbf{Acknowledgements:} The simulation was performed on resources provided by the Swedish National Infrastructure for Computing (SNIC) at HPC2N (Ume\aa).

\bibliographystyle{mnras}
\bibliography{Manuscript}

\begin{thebibliography}{}
\makeatletter
\relax
\def\mn@urlcharsother{\let\do\@makeother \do\$\do\&\do\#\do\^\do\_\do\%\do\~}
\def\mn@doi{\begingroup\mn@urlcharsother \@ifnextchar [ {\mn@doi@}
  {\mn@doi@[]}}
\def\mn@doi@[#1]#2{\def\@tempa{#1}\ifx\@tempa\@empty \href
  {http://dx.doi.org/#2} {doi:#2}\else \href {http://dx.doi.org/#2} {#1}\fi
  \endgroup}
\def\mn@eprint#1#2{\mn@eprint@#1:#2::\@nil}
\def\mn@eprint@arXiv#1{\href {http://arxiv.org/abs/#1} {{\tt arXiv:#1}}}
\def\mn@eprint@dblp#1{\href {http://dblp.uni-trier.de/rec/bibtex/#1.xml}
  {dblp:#1}}
\def\mn@eprint@#1:#2:#3:#4\@nil{\def\@tempa {#1}\def\@tempb {#2}\def\@tempc
  {#3}\ifx \@tempc \@empty \let \@tempc \@tempb \let \@tempb \@tempa \fi \ifx
  \@tempb \@empty \def\@tempb {arXiv}\fi \@ifundefined
  {mn@eprint@\@tempb}{\@tempb:\@tempc}{\expandafter \expandafter \csname
  mn@eprint@\@tempb\endcsname \expandafter{\@tempc}}}

\bibitem[\protect\citeauthoryear{Ahmed et~al.,}{Ahmed et~al.}{2013}]{Ahmed13}
Ahmed H.,  et~al., 2013, Phys. Rev. Lett., 110, 205001

\bibitem[\protect\citeauthoryear{Amatucci}{Amatucci}{1999}]{Amatucci99}
Amatucci W.~E.,  1999, J. Geophys. Res., 104, 14481

\bibitem[\protect\citeauthoryear{Arber et~al.,}{Arber et~al.}{2015}]{Arber15}
Arber T.~D.,  et~al., 2015, Plasma Phys. Controll. Fusion, 57, 113001

\bibitem[\protect\citeauthoryear{Bale, Hull, Larson, Lin, Muschietti, Kellog,
  Goetz  \& Monson}{Bale et~al.}{2002}]{Bale02}
Bale S.~D.,  Hull A.,  Larson D.~E.,  Lin R.~P.,  Muschietti L.,  Kellog P.~J.,
   Goetz K.,   Monson S.~J.,  2002, Astrophys. J., 575, L25

\bibitem[\protect\citeauthoryear{Blondin \& Marks}{Blondin \&
  Marks}{1996}]{Blondin96}
Blondin J.~M.,  Marks B.~S.,  1996, New Astron., 1, 235

\bibitem[\protect\citeauthoryear{Blondin, Chevalier  \& Frierson}{Blondin
  et~al.}{2001}]{Blondin91}
Blondin J.~M.,  Chevalier R.~A.,   Frierson D.~M.,  2001, Astrophys. J., 563,
  806

\bibitem[\protect\citeauthoryear{Bret}{Bret}{2015}]{Bret15}
Bret A.,  2015, J. Plasma Phys., 81, 455810202

\bibitem[\protect\citeauthoryear{Chevalier, Blondin  \& Emmering}{Chevalier
  et~al.}{1992}]{Chevalier92}
Chevalier R.~A.,  Blondin J.~M.,   Emmering R.~T.,  1992, Astrophys. J., 392,
  118

\bibitem[\protect\citeauthoryear{Dieckmann, Ynnerman, Chapman, Rowlands  \&
  Andersson}{Dieckmann et~al.}{2004}]{Dieckmann04b}
Dieckmann M.~E.,  Ynnerman A.,  Chapman S.~C.,  Rowlands G.,   Andersson N.,
  2004, \mn@doi [Phys. Scripta] {10.1238/Physica.Regular.069a00456}, 69, 456

\bibitem[\protect\citeauthoryear{Dieckmann, Meli, Shukla, Drury  \&
  Mastichiadis}{Dieckmann et~al.}{2008}]{Dieckmann08}
Dieckmann M.~E.,  Meli A.,  Shukla P.~K.,  Drury L. O.~C.,   Mastichiadis A.,
  2008, Plasma Phys. Controll. Fusion, 562, 065020

\bibitem[\protect\citeauthoryear{Dieckmann, Ahmed, Sarri, Doria, Kourakis,
  Romagnani, Pohl  \& Borghesi}{Dieckmann et~al.}{2013a}]{Dieckmann13}
Dieckmann M.~E.,  Ahmed H.,  Sarri G.,  Doria D.,  Kourakis I.,  Romagnani L.,
  Pohl M.,   Borghesi M.,  2013a, Phys. Plasmas, 20, 042111

\bibitem[\protect\citeauthoryear{Dieckmann, Sarri, Doria, Pohl  \&
  Borghesi}{Dieckmann et~al.}{2013b}]{Dieckmann14}
Dieckmann M.~E.,  Sarri G.,  Doria D.,  Pohl M.,   Borghesi M.,  2013b, Phys.
  Plasmas, 20, 102112

\bibitem[\protect\citeauthoryear{Dieckmann, Sarri, Doria, Ahmed  \&
  Borghesi}{Dieckmann et~al.}{2015a}]{Dieckmann15b}
Dieckmann M.~E.,  Sarri G.,  Doria D.,  Ahmed H.,   Borghesi M.,  2015a, New J.
  Phys., 16, 073001

\bibitem[\protect\citeauthoryear{Dieckmann, Bock, Ahmed, Doria, Sarri, Ynnerman
   \& Borghesi}{Dieckmann et~al.}{2015b}]{MagField15}
Dieckmann M.~E.,  Bock A.,  Ahmed H.,  Doria D.,  Sarri G.,  Ynnerman A.,
  Borghesi M.,  2015b, Phys. Plasmas, 22, 072104

\bibitem[\protect\citeauthoryear{Dieckmann et~al.,}{Dieckmann
  et~al.}{2015c}]{Dieckmann15}
Dieckmann M.~E.,  et~al., 2015c, Phys. Rev. E, 92, 031101

\bibitem[\protect\citeauthoryear{Dum}{Dum}{1978}]{Dum78}
Dum C.~T.,  1978, Phys. Fluids, 21, 945

\bibitem[\protect\citeauthoryear{Edens, Adams, Rambo, Ruggles, Smith, Porter
  \& Ditmire}{Edens et~al.}{2010}]{Edens10}
Edens A.~D.,  Adams R.~G.,  Rambo P.,  Ruggles L.,  Smith I.~C.,  Porter J.~L.,
    Ditmire T.,  2010, Phys. Plasmas, 17, 112104

\bibitem[\protect\citeauthoryear{Folini \& Walder}{Folini \&
  Walder}{2006}]{Folini06}
Folini D.,  Walder R.,  2006, Astron. Astrophys., 459, 1

\bibitem[\protect\citeauthoryear{Folini, Walder  \& Favre}{Folini
  et~al.}{2014}]{Folini14}
Folini D.,  Walder R.,   Favre J.~M.,  2014, Astron. Astrophys., 562, A112

\bibitem[\protect\citeauthoryear{Forslund \& Freidberg}{Forslund \&
  Freidberg}{1971}]{Forslund71}
Forslund D.~W.,  Freidberg J.~P.,  1971, Phys. Rev. Lett., 27, 1189

\bibitem[\protect\citeauthoryear{Forslund \& Shonk}{Forslund \&
  Shonk}{1970a}]{Forslund70}
Forslund D.~W.,  Shonk C.~R.,  1970a, Phys. Rev. Lett., 25, 281

\bibitem[\protect\citeauthoryear{Forslund \& Shonk}{Forslund \&
  Shonk}{1970b}]{Forslund70b}
Forslund D.~W.,  Shonk C.~R.,  1970b, Phys. Rev. Lett., 25, 1699

\bibitem[\protect\citeauthoryear{Gamezo, Khokhlov, Oran, Chtchelkanova  \&
  Rosenberg}{Gamezo et~al.}{2003}]{Gamezo03}
Gamezo V.~N.,  Khokhlov A.~M.,  Oran E.~S.,  Chtchelkanova A.~Y.,   Rosenberg
  R.~O.,  2003, Sci., 299, 77

\bibitem[\protect\citeauthoryear{Hansen, Edwards, Froula, Edens, Gregori  \&
  Ditmire}{Hansen et~al.}{2006}]{Hansen06}
Hansen J.~F.,  Edwards M.~J.,  Froula D.~H.,  Edens A.~D.,  Gregori G.,
  Ditmire T.,  2006, Phys. Plasmas, 13, 112101

\bibitem[\protect\citeauthoryear{Heitsch, Slyz, Devriendt, Hartmann  \&
  Burkert}{Heitsch et~al.}{2007}]{Heitsch07}
Heitsch F.,  Slyz A.~D.,  Devriendt J. E.~G.,  Hartmann L.~W.,   Burkert A.,
  2007, Astrophys. J., 665, 445

\bibitem[\protect\citeauthoryear{Hershkowitz}{Hershkowitz}{1981}]{Hershkowitz81}
Hershkowitz N.,  1981, J. Geophys. Res., 86, 3307

\bibitem[\protect\citeauthoryear{Karimabadi, Omidi  \& Quest}{Karimabadi
  et~al.}{1991}]{Karimabadi91}
Karimabadi H.,  Omidi N.,   Quest K.~B.,  1991, Geophys. Res. Lett., 18, 1813

\bibitem[\protect\citeauthoryear{Kato \& Takabe}{Kato \& Takabe}{2010}]{Kato10}
Kato T.~N.,  Takabe H.,  2010, Phys. Plasmas, 17, 032114

\bibitem[\protect\citeauthoryear{Lamberts, Fromang  \& Dubus}{Lamberts
  et~al.}{2011}]{Lamberts11}
Lamberts A.,  Fromang S.,   Dubus G.,  2011, Mon. Not. R. Astron. Soc., 418,
  2618

\bibitem[\protect\citeauthoryear{McLeod \& Whitworth}{McLeod \&
  Whitworth}{2013}]{McLeod13}
McLeod A.~D.,  Whitworth A.~P.,  2013, Mon. Not. R. Astron. Soc., 431, 710

\bibitem[\protect\citeauthoryear{Michaut, Cavet, Bouquet, Roy  \&
  Nguyen}{Michaut et~al.}{2012}]{Bouquet2}
Michaut C.,  Cavet C.,  Bouquet S.~E.,  Roy F.,   Nguyen H.~C.,  2012,
  Astrophys. J., 759, 78

\bibitem[\protect\citeauthoryear{Palotti, Heitsch, Zweibel  \& Huang}{Palotti
  et~al.}{2008}]{Palotti08}
Palotti M.~L.,  Heitsch F.,  Zweibel E.~G.,   Huang Y.~M.,  2008, Astrophys.
  J., 678, 234

\bibitem[\protect\citeauthoryear{Piriz, Cortazar, Cela  \& Tahir}{Piriz
  et~al.}{2006}]{Piriz06}
Piriz A.~R.,  Cortazar O.~D.,  Cela J. J.~L.,   Tahir N.~A.,  2006, Am. J.
  Phys., 74, 1095

\bibitem[\protect\citeauthoryear{Porth, Komissarov  \& Keppens}{Porth
  et~al.}{2014}]{Porth14}
Porth O.,  Komissarov S.~S.,   Keppens R.,  2014, Month. Not. R. Astron. Soc.,
  443, 547

\bibitem[\protect\citeauthoryear{Raadu}{Raadu}{1989}]{Raadu89}
Raadu M.~A.,  1989, Phys. Rep., 178, 25

\bibitem[\protect\citeauthoryear{Sanz, Bouquet  \& Murakami}{Sanz
  et~al.}{2011}]{Bouquet1}
Sanz J.,  Bouquet S.,   Murakami M.,  2011, Astrophys. Space Sci., 336, 195

\bibitem[\protect\citeauthoryear{Sharp}{Sharp}{1984}]{Sharp84}
Sharp D.~H.,  1984, Phys. D, 12, 3

\bibitem[\protect\citeauthoryear{Stawarz \& Ostrowski}{Stawarz \&
  Ostrowski}{2002}]{Stawarz02}
Stawarz L.,  Ostrowski M.,  2002, Astrophys. J., 578, 763

\bibitem[\protect\citeauthoryear{Vishniac}{Vishniac}{1983}]{Vishniac83}
Vishniac E.~T.,  1983, Astrophys. J, 1, 152

\bibitem[\protect\citeauthoryear{Vishniac}{Vishniac}{1994}]{Vishniac94}
Vishniac E.~T.,  1994, Astrophys. J., 428, 186

\bibitem[\protect\citeauthoryear{Walder \& Folini}{Walder \&
  Folini}{2000}]{Walder00}
Walder R.,  Folini D.,  2000, Astrophys. Space Sci., 274, 343

\bibitem[\protect\citeauthoryear{van Marle \& Keppens}{van Marle \&
  Keppens}{2012}]{vanMarle1}
van Marle A.~J.,  Keppens R.,  2012, Astron. Astrophys., 547, A3

\bibitem[\protect\citeauthoryear{van Marle, Keppens  \& Meliani}{van Marle
  et~al.}{2011}]{vanMarle2}
van Marle A.~J.,  Keppens R.,   Meliani Z.,  2011, Astron. Astrophys., 527, A3

\makeatother
\end{thebibliography}
\label{lastpage}
\end{document}